\documentclass[epj]{svjour}
\usepackage{amsmath,amssymb,amsfonts}

\begin{document}

\title{Comment on ``A local realist model for correlations of the
  singlet state'' (De Raedt \emph{et al}., Eur.\ Phys.\ J.\ B
  \textbf{53}: 139-142, 2006)}
\titlerunning{Comment on ``A local realist model\ldots}
\author{M.P.  Seevinck\inst{1} \thanks{E-mail:
    {\tt{seevinck@phys.uu.nl}} } \and J.-\AA. Larsson\inst{2}
  \thanks{E-mail: {\tt{jalar@mai.liu.se}} } }


\institute{Institute of History and Foundations of Science, Utrecht
  University, P.O Box 80.000, 3508 TA Utrecht, The Netherlands.  \and
  Matematiska Institutionen, Link\"opings Universitet, SE-581 83
  Link\"oping, Sweden.}

\date{Received: date / Revised version: date}

\abstract{De Raedt \emph{et al.}\ (Eur.\ Phys.\ J.\ B {\bf
    53}: 139-142, 2006) have claimed to provide a 
  local realist model for correlations of the singlet state in the
  familiar Einstein-Podolsky-Rosen-Bohm (EPRB) experiment when
  time-coincidence is used to decide which detection events should
  count in the analysis, and furthermore that this suggests that it is
  possible to construct local realistic models that can reproduce the
  quantum mechanical expectation values.  In this letter we show that
  these conclusions cannot be upheld since their model exploits the
  so-called coincidence-time loophole. When this is properly taken into account no startling conclusions can be drawn about local realist modelling of quantum
  mechanics.
} 

\maketitle
\noindent
De Raedt \emph{et al.} \cite{deraedt} have recently claimed to have
constructed a local realist model for correlations of the singlet
state, in which time-coincidence is used to decide which detection
events should count in the analysis.  Furthermore, they claim that
their model maximally violates the well-known
Clauser-Horne-Shimony-Holt (CHSH) inequality \cite{CHSH}, and conclude
that their work ``suggests that it is possible to construct
event-based simulation models that satisfy Einstein's criteria of
local causality and realism and can reproduce the expectation values
calculated by quantum theory''\cite{deraedt}.

Here, we will put the model used by de Raedt \emph{et al.} in its
proper context and show that, although the model gives the sinusoidal
correlations familiar from quantum mechanics, the conclusions drawn by
De Raedt \emph{et al.} cannot be upheld. This is because their model
is based on post-selection, something which is known to enable
quantum-like correlations from a local realist model. This possibility
was first noted by Pearle in the late sixties \cite{pearle70} and has
received quite some interest in the years that followed
\cite{ClauHorn,GargMerm} and has been especially active more recently
\cite[\ldots]{jalar98a,jalar99a,gisin99,eberhard93,jalar00c,%
massarpironio,cabellolarsson}; a full list of references would be
immense. We will see below that the model in its general form exploits
the so-called ``coincidence-time loophole'' \cite{larssongill}, and
that the usual CHSH inequality is inappropriate for this situation
because of the postselection. The appropriately modified inequality is
not violated by the de Raedt \emph{et al.}  model, even though it
gives quantum-like correlations.  Notable is also that the
post-selection procedure for coincidence in time used by De Raedt
\emph{et al.} is not used in all experimental realizations of the EPRB
experiment \cite{EPR,Bohm51} and consequently their model cannot
reproduce all experimental realizations of the EPRB experiment (see
e.g., \cite{rowe01}).  Since this is the case, we would argue that no
startling conclusion can be drawn from the model about local realist
modelling of quantum-mechanical singlet correlations, nor about local
realist modelling of quantum mechanics in general.

Let us first go through some notation. The setup of the EPRB
experiment has two measurement stations $i=1,2$ with Stern-Gerlach
magnets that can be set to measurement directions $\vec{a}_1$ and
$\vec{a}_2$ respectively, and the angular difference between these
settings is denoted $\alpha$. The local hidden variable in the model
of De Raedt \emph{et al.} is denoted $S_{n,i}$ (event number $n$,
station $i$) and is a direction in space.  From this hidden variable,
the model gives results $x_{n,i}$ and detection times $t_{n,i}$ with
time resolution $\tau$.  Coincidences only occur when the detection
times are within a time window of width $W$, i.e., when
\begin{equation}
  |t_{n,1}-t_{n,2}|\le W.  
\end{equation}
These coincidences are used to calculate the correlations of the
outcomes $x_{n,i}$ for different setting combinations (see Eq.\
(\cite{deraedt}:3)).  This is exactly the same kind of modeling as
that discussed in \cite{larssongill}.

In the limit where $W=\tau\rightarrow 0$ (i.e., where the time window
and the time-tag resolution both go to zero) De Raedt \emph{et al.}
obtain the sinusoidal correlation of the singlet state from their
model. They then argue that this correlation violates the well-known
CHSH inequality
\begin{equation}
  \big|E(\vec{a},\vec{c})-E(\vec{a},\vec{d}) 
  +E(\vec{b},\vec{c})+E(\vec{b},\vec{d})\big| \leq 2, 
  \label{eq:1}
\end{equation} 
where $E(\vec{a},\vec{b})$ is the correlation between outcomes for
settings $\vec{a}$ and $\vec{b}$, etc. They furthermore claim that the
maximal quantum violation is obtained using their model (i.e., where
the left-hand side of (\ref{eq:1}) is $2\sqrt{2}$). However,
inequality (\ref{eq:1}) is not valid for their model; the correlations
they calculate (that reach $2\sqrt2$) are not on the form of the ones
on the left-hand side of (\ref{eq:1}).

The problem is the postselection of events that are close enough in
time for which ``[\ldots] the simultaneity of two detection events
will depend on \emph{both settings}, even though the underlying
physical processes that control this are completely local''
\cite{larssongill}. A postselection procedure of this kind invalidates
the original CHSH inequality (\ref{eq:1}) since the ensemble on which
the correlations are evaluated changes with the settings, while
(\ref{eq:1}) requires them to remain the same; see
\cite{jalar98a,larssongill}. The correlation calculated in
\cite{deraedt} is not $E(\vec{a}_1,\vec{a}_2)$, as was claimed, but
\begin{equation}
  E(\vec{a}_1,\vec{a}_2|\Lambda_{\vec{a}_1\vec{a}_2})=-\vec{a}_1\cdot\vec{a}_2,
  \label{eq:35}
\end{equation}
where $E(\vec{a}_1,\vec{a}_2|\Lambda_{\vec{a}_1\vec{a}_2})$ is the
\emph{conditional} correlation, conditioned on a coincidence for the
settings $\vec{a}_1$ and $\vec{a}_2$. Consequently, inequality
(\ref{eq:1}) cannot be used. The appropriate inequality includes this
conditioning on coincidence and reads \cite{larssongill}
\begin{equation}
  \begin{split}
    \big|E&(\vec{a},\vec{c}|\Lambda_{\vec{a}\vec{c}})-E(\vec{a},\vec{d}|
    \Lambda_{\vec{a}\vec{d}}) \\&+
    E(\vec{b},\vec{c}|\Lambda_{\vec{b}\vec{c}})+
    E(\vec{b},\vec{d}|\Lambda_{\vec{b}\vec{d}})\big|\leq
    \frac{6}{\gamma} -4.
  \end{split}
 \label{belluseful}
\end{equation}
The quantity $\gamma$ is the probability of coincidence. Quantum
mechanical correlations that violate the CHSH inequality (\ref{eq:1})
by the value $2\sqrt2$ will violate (\ref{belluseful}) only if
$\gamma>\gamma_{0}=3-3/\sqrt{2}\approx 0.8787$; this bound is
necessary and sufficient (see \cite{larssongill} for further details).
That is, if $\gamma\le\gamma_0$, it is possible to construct a local
realist model that gives a value of $2\sqrt2$ on the left-hand side.
Such a model is given in \cite{larssongill} which furthermore
saturates the bound.

Let us now go back to the model of De Raedt \emph{et al}. If
$W=\tau\ll1$, we have
\begin{subequations}
  \begin{align}
    \gamma&\le8\tau\cot\frac\alpha2,&\alpha\neq0,\\
    \gamma&\lesssim 6\pi\tau^{2/3},&\alpha=0,
  \end{align}
\end{subequations}
(see Appendix A) so evidently the value of $\gamma$ approaches zero
when $W=\tau\to0$.  This is below the bound specified above and,
although the model gives sinusoidal correlations --- and may be
interesting as such --- it does not violate the relevant Bell inequality
(\ref{belluseful})\footnote{Actually, since De Raedt \emph{et al.} use
  $W=\tau$ (time window length equals time resolution), they are in
  effect using the more well-studied ``efficiency loophole''
  \cite{pearle70,GargMerm,jalar98a}, but that is perhaps more of a
  technical side note.}.  For other local realist models with a
sinusoidal interference pattern, but with a nonzero probability of
coincidence see, e.g., Refs.~\cite{jalar99a,gisin99}.

In conclusion, even though the model by De Raedt \emph{et
  al}.\cite{deraedt} does give (conditional) correlations as strong as
quantum mechanics, it does not model the singlet state as such,
because in the model the probability of coincidence must go to zero to
obtain the sinusoidal interference pattern. This means that the model
does not violate the relevant Bell inequality, because it is far below
the relevant coincidence-probability bound $\gamma_0\approx0.8787$.
In addition, this is far below the coincidence probabilities of
previously published local realist models \cite{jalar99a,gisin99}.
Finally, even though De Raedt \emph{et al.}  claim their model can
reproduce the coincidences of recent experimental results, it cannot:
optical experiments reach $\gamma\approx0.05$ \cite{WJSWZ} and
ion-trap experiments even reach $\gamma=1$ \cite{rowe01}; the latter
consequently does not fall prey to the coincidence-time
loophole.\footnote{Nor to the detection loophole.}  This reinforces
the conclusion drawn in \cite{larssongill} of the importance of
eliminating postselection in future EPRB experiments, and, as we've
seen here, postselection must be duly accounted for in any local
realist modelling of them.

\section*{Appendix A}
\noindent
The probability of coincidences $\gamma$ of the model in
\cite{deraedt} is given by the denominator of (\cite{deraedt}:6) 
and can be calculated using the density of coincidences
\begin{equation}
  P(T_1,T_2,W)\le\tau\frac{\min(T_1,T_2)}{T_1T_2}.
\label{eq:22}
\end{equation}
The above bound is valid when $W=\tau$ and is given in \cite{deraedt}.
One should also remember that 
\begin{equation}
  P(T_1,T_2,W)\le 1,
  \label{eq:6}
\end{equation}
since it is the probability that a coincidence occurs given the values
of the hidden variables $S_{n,i}$ and the settings $\vec{a}_i$. When
$\tau\ll1$ and $\alpha\neq0$ (i.e., $\vec{a}_1\neq\vec{a}_2$) the
inequality (\ref{eq:6}) is automatically satisfied when (\ref{eq:22})
is, and
\begin{equation}
  \label{eq:2}
  \begin{split}
    \gamma&=\int_0^\pi\int_0^{2\pi}P(T_1,T_2,W)\sin\theta d\theta d\phi\\
    &\le \tau
    \int_0^\pi\int_0^{2\pi}\frac{\min(T_1,T_2)}{T_1T_2}\sin\theta
    d\theta d\phi\\
    &= 2\tau\int_0^{2\pi}
    \frac{\min\left(\sin^2\phi,\sin^2(\phi-\alpha)\right)}
    {\sin^2\phi\sin^2(\phi-\alpha)} d\phi
    =8\tau\cot\frac\alpha2.
 \end{split}
\end{equation}
For the case when $\alpha=0$ (i.e., $\vec{a}_1=\vec{a}_2$ and
$T_1=T_2$), the requirement (\ref{eq:6}) needs to be taken into
account, and
\begin{align}
  \label{eq:5}
  \notag
  \gamma&=\int_0^\pi\int_0^{2\pi}P(T_1,T_2,W)\sin\theta d\theta d\phi\\
  \notag &\le
  \int_0^\pi\int_0^{2\pi}\min\left(\dfrac{\tau}{T_i},1\right)\sin\theta
  d\theta d\phi\\
  \notag &=
  \int_0^\pi\int_0^{2\pi}\min\left(
    \dfrac{\tau}{(1-\cos^2\phi\sin^2\theta)^{3/2}},1\right)\sin\theta
  d\theta d\phi\\
  \notag &=
  4\pi\left(\tau^{2/3}\sqrt{1-\tau^{2/3}}
    +\frac{\tau^{2/3}}{1+\sqrt{1-\tau^{2/3}}}\right)\\
  &\approx6\pi\tau^{2/3}.
\end{align}
The last approximation is good when $\tau\ll1$.



\end{document}